\documentclass[a4paper,11pt]{article}
\usepackage[width=15cm,height=23cm]{geometry} 
\usepackage[usenames]{color}
\usepackage{amsfonts,amssymb}
\usepackage{theorem}
\usepackage[dvips]{graphicx}

\long\def\symbolfootnote[#1]#2{\begingroup%
\def\thefootnote{\fnsymbol{footnote}}\footnote[#1]{#2}\endgroup}

\def\slshp{p\!\!\!\slash}

\begin{document}

\vskip 1.0 truecm
\rightline{FR-PHENO-032}

\vskip 2.0 truecm
\Large
\bf
\centerline{Resummed corrections to the $\rho$ parameter}
\centerline{in the complex mass scheme}
\normalsize \rm

\large
\rm
\vskip 1.3 truecm
\centerline{D.~Bettinelli
\footnote{e-mail: {\tt daniele.bettinelli@physik.uni-freiburg.de}}, 
J.~J.~van der Bij
\footnote{e-mail: {\tt jochum@physik.uni-freiburg.de}}}

\normalsize
\medskip
\begin{center}
Physikalisches Institut, Albert-Ludwigs-Universit\"at Freiburg\\
Hermann-Herder-Str. 3, D-79104 Freiburg im Breisgau, Germany.
\end{center}

\vskip 0.7  truecm
\normalsize
\bf
\centerline{Abstract}

\rm

\begin{quotation}
We present all order results for the heavy top corrections to the $\rho$ parameter in the complex mass scheme.
We derive translation formulas between the complex mass and the on-shell scheme  
and show that they are ultimately equivalent.
We show that a naive treatment with a fixed width for the top quark cannot give even approximately correct results. 
\end{quotation}

\newpage

\section{Introduction}
\label{sec.intro}
%

In a recent paper \cite{Bettinelli:2010gm} the $SU(N_F) \times U(1)$ model at the leading order in the 
large $N_F$-limit has been used in  order to compute the exact leading top quark contribution to the 
$\rho$ parameter and its perturbative expansion to all orders in the interaction strength.  
This paper improved earlier calculations \cite{Aoki} along similar lines. 

In the $SU(N_F) \times U(1)$ model the exact top quark propagator, at the leading order in $N_F$, 
can be obtained simply by resumming one-loop self-energy insertions, thereby including the finite width effects 
due to the fact that the top quark is an unstable particle. 

The goal of this Letter is to compare the results obtained in Ref.~\cite{Bettinelli:2010gm}, by 
adopting the on-shell scheme in order to renormalize the self-energy insertions, with those 
that one gets within the framework of the complex mass scheme \cite{complex}.
The complex mass scheme has been developed in order to provide a gauge invariant treatment 
of resonances \cite{Beenakker:1996uv} and consists in choosing the whole complex pole 
 of the resummed propagator (instead of its real part only) as the renormalization point. 

At the perturbative level the two renormalization schemes can be compared 
by expanding the coefficients of the complex subtraction point in powers of the 
interaction strength of the theory in the on-shell scheme.  
The simplified perturbative framework provided by the $SU(N_F) \times U(1)$ model at the leading order in $N_F$
can be successfully used as a testing ground for the predictions on measurable quantities in different 
 renormalization schemes. 
In particular the leading top contribution to the $\rho$ parameter 
 has been computed to all orders in perturbation theory both in the on-shell \cite{Bettinelli:2010gm} 
and in the complex mass scheme.
As we demonstrate in this Letter, it turns out that the two results coincide, order by order in perturbation theory, 
once they are expressed  as series in powers of the same interaction strength.

Finally, the independence from the chosen renormalization scheme can be proven also beyond the 
 perturbative approximation. In this connection the prescription of subtracting minimally the tachyon pole 
 from the resummed propagator \cite{Bettinelli:2010gm, Binoth:1997pd} can be used both in the on-shell 
and in the complex mass scheme in order to find a tachyon-free representation of the top propagator. 
The latter allows us to determine the exact leading top contribution to the $\rho$ parameter, thereby
 showing the equality of the two renormalization schemes.

\section{Top quark self-energy}
\label{sec.top}
%

In this Section we shall discuss the renormalization of the top quark self-energy in the complex mass scheme
 and derive the exact top propagator at the leading order in the large $N_F$ limit. Moreover, a comparison with the 
 on-shell scheme will be presented. Here and in the subsequent Sections, we will follow the notation 
 introduced in Ref.~\cite{Bettinelli:2010gm}.

The exact bare top quark self-energy at the leading order in the large $N_F$ limit is given by
\begin{eqnarray}
\Sigma_t(p) = \frac{\sqrt{2}}{16 \pi^2}\,G_F\, m^2_{t,0}\,N_F
\Big[\frac{2}{D-4}+\log\Big(\!\!-\frac{p^2}{\Lambda_B^2}-i\epsilon\Big)\Big]\, \slshp\,\omega_+ \,,
\label{eq.osr.1}
\end{eqnarray}
where $m_{t,0}$ is the bare top quark mass, while $\Lambda_B$ is a regulator-dependent 
quantity with the dimension of a mass whose explicit expression is not needed in the subsequent analysis.

The natural way to incorporate the finite width effects of an unstable 
 particle is the resummation of the corresponding self-energy 
insertions. This leads us to consider the Dyson resummed top quark 
 propagator instead of the Born one. 
 We report here only the component of the resummed 
 propagator  with positive chirality because it is the only one which 
 gives a non-vanishing contribution to the $\rho$ parameter.
\begin{eqnarray}
&&
D_t(p) = \frac{i\,\slshp \, a_0(p^2)\,\omega_+}
{a_0(p^2)\,p^2-m^2_{t,0} + i \epsilon}\,, ~~~ {\rm where} \nonumber\\
&& a_0(p^2) = 1 - \frac{\sqrt{2}}{16 \pi^2}\,G_F\, m^2_{t,0}\,N_F
\Big[\frac{2}{D-4}+\log\Big(\!\!-\frac{p^2}{\Lambda_B^2}-i\epsilon\Big)\Big]\,.
\label{eq.new.1}
\end{eqnarray}
In the on-shell renormalization scheme one requires that the real part of the 
denominator of the resummed propagator in the above equation vanishes 
when evaluated at a real subtraction point, $p^2 = m^2_t$. This allows us to express the bare mass 
 of the top quark in terms of the subtraction point $m_t$  
\begin{eqnarray}
&&
m^2_{t,0} = {\rm Re}\big[ a_0(m_t^2)\big]\, m_t^2 = m_t^2 - \alpha_t\,
\Big[\frac{2}{D-4}+\log\Big(\frac{m_t^2}{\Lambda_B^2}\Big)\Big]\,m_t^2 \,,\nonumber\\
&&
{\rm where}~~ \alpha_t = \frac{\sqrt{2}}{16 \pi^2}\,G_F\, m^2_{t}\,N_F \,.
\label{eq.new.2}
\end{eqnarray}

By substituting the above equation into eq.(\ref{eq.new.1}), we obtain the 
 on-shell renormalized top quark propagator at the leading order in the large-$N_F$ limit
\begin{eqnarray}
\widehat{D}_t(p) = \frac{i\,\slshp \, a(p^2)\,\omega_+}
{a(p^2)\,p^2-m^2_t+i \epsilon}\,,~~ {\rm where}~~
a(p^2) = 1- \alpha_t\, \log\Big(\!\!-\frac{p^2}{m^2_t}-i\epsilon\Big)\,.
\label{eq.osr.4bis}
\end{eqnarray}

In the complex mass scheme (see for instance Ref.~\cite{Beenakker:1996uv}) one introduces 
a complex subtraction point, namely $\mu^2_t = m^2_{t,c}\big(1-i\,w_t\big)$, 
and requires that both the real and the imaginary part of the denominator of the resummed 
top propagator (\ref{eq.new.1}) vanish when computed at $p^2 = \mu_t^2$. This gives us the following conditions
\begin{eqnarray}
m^2_{t,0} = {\rm Re}\big[ a_0(\mu_t^2)\, \mu_t^2\big]\,, ~~
{\rm Im}\big[ a_0(\mu_t^2)\, \mu_t^2\big] = 0\,. 
\label{eq.new.5}
\end{eqnarray}

In order to evaluate the top quark self-energy at a complex square momentum, we perform the appropriate expansion about
 the real part of the complex subtraction point
\begin{eqnarray}
a_0(\mu_t^2) = a_0(m_{t,c}^2)+\sum^\infty_{j=1} \frac{\big(\mu_t^2-m_{t,c}^2\big)^j}{j!}\,
a_0^{(j)}(p^2)\arrowvert_{p^2=m_{t,c}^2}\,,
\label{eq.cms.5}
\end{eqnarray}
where $a_0^{(j)}(p^2)$ is the $j$-th derivative of $a_0(p^2)$ w.r.t. its argument.
By using the definition of $a_0(p^2)$ in the second line of eq.(\ref{eq.new.1}) 
 and the above expansion, one can easily get
\begin{eqnarray}
&&\!\!\!\!\!\!\!\!\!\!\!\!\!\!
a_0(\mu_t^2) = 1-\frac{\sqrt{2}}{16 \pi^2}\,G_F\, m^2_{t,0}\,N_F \,\Big[\frac{2}{D-4}+
\log\Big(\frac{m_{t,c}^2}{\Lambda_B^2}\Big)+\frac{1}{2}\,\log\Big(1+w_t^2\Big)\nonumber\\
&&~~~~~~~~~~~~~~~~~~~~~~~~~~~~~~~~~
-i\pi-i\,\arctan w_t \Big]\,.
\label{eq.cms.7}
\end{eqnarray}
The bare top quark mass in the complex mass scheme reads
\begin{eqnarray}
&&\!\!\!\!\!\!\!
m^2_{t,0} = m^2_{t,c}-\alpha_{t,c}\Big[\frac{2}{D-4}+
\log\Big(\frac{m_{t,c}^2}{\Lambda_B^2}\Big) - W \Big]\, m_{t,c}^2\,,~{\rm where}\nonumber\\
&&\!\!\!\!\!\!\!
\alpha_{t,c} = \frac{\sqrt{2}}{16 \pi^2}\,G_F\, m^2_{t,c}\,N_F\,, ~ {\rm while ~ with}\nonumber\\
&&\!\!\!\!\!\!\!
W = -\frac{1}{2}\log\Big(1+w_t^2\Big)+\pi w_t + w_t \arctan w_t \,,
\label{eq.cms.6}
\end{eqnarray}
we denote the finite width effects that stem from the requirement of subtracting the whole complex pole 
instead of its real part.  

By substituting the bare top quark mass given in the first line of 
eq.(\ref{eq.cms.6}) into the Dyson propagator (\ref{eq.new.1}), one 
obtains the renormalized top quark propagator in the complex mass scheme
\begin{eqnarray}
&&\!\!\!\!\!\!\!\!\!\!
\widehat{D}_{t,c}(p) = \frac{i\,\slshp\,a_c(p^2)\,\omega_+}
{a_c(p^2)\,p^2 - m^2_{t,c} + i \epsilon}\,,~{\rm where}\nonumber\\
&&\!\!\!\!\!\!\!\!\!\! 
a_c(p^2) = 1- \alpha_{t,c} \Big[\log\Big(\!\!-\frac{p^2}{m^2_{t,c}}-i\epsilon\Big) + W \Big] \,.
\label{eq.new.6}
\end{eqnarray}
In order to determine the imaginary part of the subtraction point $\mu_t$, we impose that the imaginary part 
of the denominator of the above propagator vanishes when evaluated at $p^2 = m^2_{t,c}\big(1-i\, w_t\big)$.
By doing this, we get the following nonlinear equation 
\begin{eqnarray}
w_t-\alpha_{t,c}\big(1+w_t^2\big) 
\big(\pi  +  \arctan w_t\big) = 0 \,,
\label{eq.new.7}
\end{eqnarray}
which will allow us to express $w_t$ directly in terms of physical renormalized quantities.

At the perturbative level, the comparison between the on-shell and the complex mass scheme can be achieved by 
expanding the coefficients of the complex subtraction point, $m^2_{t,c}$ and $w_t$, in powers of the 
interaction strength of the theory in the on-shell scheme, $\alpha_t$.
Clearly, at the leading order in $\alpha_t$, one should recover the subtraction point in the on-shell scheme, thus
 $m^2_{t,c} = m^2_t + O(\alpha_t)$ and $w_t = O(\alpha_t)$. The first coefficient in the expansion of the imaginary 
 part of the complex subtraction point can be obtained by solving the linearized version of eq.(\ref{eq.new.7}), 
 which gives $w_t = \pi\, \alpha_t + O\big(\alpha_t^2\big)$. Therefore we postulate the following perturbative 
 expansions  
\begin{eqnarray}
&&
m^2_{t,c} = m^2_t\, \sum_{j=0}^\infty a_j\, \alpha_t^j\,, ~\, {\rm with}~~ a_0 = 1\,, ~~ {\rm thus} ~~
\alpha_{t,c} = \alpha_t \, \sum_{j=0}^\infty a_j\, \alpha_t^j\,, 
\nonumber\\
&&
w_t = \pi \,\alpha_t \, \sum_{j=0}^\infty b_j\, \alpha_t^j\,,~~ {\rm with}~~ b_0 = 1\,.
\label{eq.new.8}
\end{eqnarray}
If we substitute the expansions in the above equation 
into eq.(\ref{eq.new.7}),  we can solve the latter in the sense of formal power series. 
This allows us to express the coefficients  $\{b_j\}$ in terms of the expansion coefficients of $m^2_{t,c}$. 
We report here only the first few of them, since the general expression is rather cumbersome.
\begin{eqnarray}
&&
b_1 = 1+a_1\,,\nonumber\\
&&
b_2 = 1+\pi^2+ 2 a_1 + a_2\,,\nonumber\\
&&
b_3 = 1+\frac{11}{3}\pi^2+ 3 a_1\big(1+\pi^2\big) + a_1^2 + 2 a_2 +a_3 \,,
\nonumber\\&&
b_4 = 1+\frac{26}{3}\pi^2+2\pi^4+ 4 a_1\Big(1+\frac{11}{3}\pi^2\Big) + 3 a_1^2\big(1+\pi^2\big) 
+2 a_1 a_2 \nonumber\\
&&~~~~\,\, + 3 a_2\big(1+\pi^2\big) + 2 a_3+a_4 \,.
\label{eq.new.10}
\end{eqnarray}

In order to determine the coefficients $\{a_j\}$ we impose that the real part of the denominator of the 
 propagator in eq.(\ref{eq.new.6}), that is
\begin{eqnarray}
p^2- \alpha_{t,c} \Big[\log\Big(\frac{p^2}{m^2_{t,c}}\Big) + W \Big]\,p^2-m^2_{t,c}\,,
\label{eq.new.11}
\end{eqnarray}
vanishes to all orders in $\alpha_t$ when evaluated at $p^2 = m^2_t$
\footnote{Notice that the same condition, but with $p^2 = m^2_{t,c}$, 
cannot be imposed, since it is equivalent to the requirement $W = 0$ and there is no possible choice of 
the coefficients $\{a_j\}$ for which $W$ vanishes identically.}.
The first expansion coefficients turn out to be
\begin{eqnarray}
&&
a_1 = 0\,,\nonumber\\
&&
a_2 = -\pi^2 \,,\nonumber\\
&&
a_3 =-\frac{5}{2}\,\pi^2 \,,\nonumber\\
&&
a_4 = \pi^2\Big(\pi^2-\frac{9}{2}\Big) \,.
\label{eq.new.12}
\end{eqnarray}
If we substitute the above results into eq.(\ref{eq.new.10}), we get
\begin{eqnarray}
&&
b_1 = 1\,,\nonumber\\
&&
b_2 = 1\,,\nonumber\\
&&
b_3 = 1-\frac{5}{6}\pi^2\,,\nonumber\\
&&
b_4 = 1-\frac{23}{6}\pi^2\,.
\label{eq.new.13}
\end{eqnarray}
By using eq.(\ref{eq.new.13}), we obtain the perturbative expansion of the finite width effects $W$ 
 in powers of $\alpha_t$
\begin{eqnarray}
W = \pi^2\,\alpha_t\,\Big[1+\frac{3}{2}\,\alpha_t+2 \alpha_t^2+\Big(\frac{5}{2}-\frac{11}{2}\,\pi^2\Big)\alpha_t^3+
 \big(3-5 \pi^2\big) \alpha_t^4 + O\big(\alpha_t^5\big)\Big]\,.
\label{eq.new.13bis}
\end{eqnarray}  
%

\subsection{Tachyonic regularization}
\label{sect.tac}
%

Besides the complex pole at $p^2 = \mu^2_t$, corresponding to the unstable top quark, 
 the propagator in eq.(\ref{eq.new.6}) contains a tachyon pole. Its Euclidean position, $p^2 = -\Lambda^2_{T,c}$, 
can be obtained by solving numerically the following equation
\begin{eqnarray}
1+ \frac{1}{\lambda_{T,c}^2} = \alpha_{t,c}\,\Big[\log\big(\lambda_{T,c}^2\big)+W\Big]\,,
~~{\rm where}~~ \lambda^2_{T,c} = \frac{\Lambda^2_{T,c}}{m^2_{t,c}}\,.
\label{eq.tac}
\end{eqnarray}
The tachyon pole induces causality violation effects in the theory and makes all the  Wick-rotated Feynman integrals 
 ill-defined. Following the procedure devised in Ref.~\cite{Binoth:1997pd}, we modify the top propagator in 
 eq.(\ref{eq.new.6}) by subtracting minimally from it the tachyonic pole. This leads us to the following  
tachyon-free representation of the resummed top propagator 
\begin{eqnarray}
&&
\widehat{D}_t(p) = \frac{i\,\slshp \,\omega_+}{1-\kappa_c}\Big[ \frac{a_c(p^2)}{a_c(p^2)\,p^2-m^2_{t,c}}
-\frac{\kappa_c}{p^2+\Lambda_{T,c}^2}\Big]\,, ~~{\rm where}\nonumber\\
&&
\kappa_c = \frac{1}{1+\alpha_{t,c}\, \lambda_{T,c}^2} \approx \frac{1}{\alpha_{t,c}}\, 
\exp\Big(-\frac{1}{\alpha_{t,c}}\Big)
\label{eq.kall.3}
\end{eqnarray}
is the residuum at the tachyonic pole. 
 The approximation of $\kappa_c$ in the second line of the above equation stems from the leading term 
of eq.(\ref{eq.tac}). Moreover, in the same equation, the prefactor $1/(1-\kappa_c)$ ensures the 
correct normalization of the spectral function after the subtraction of the tachyon 
(see Ref.~\cite{Bettinelli:2010gm} for further details).

In order to compare the numerical results for the tachyonic pole in the complex mass scheme with 
those obtained in the on-shell scheme \cite{Bettinelli:2010gm}, 
 we expand $\lambda_{T,c}^2$ in powers of $\alpha_t$ 
\begin{eqnarray}
\lambda^2_{T,c} = \lambda^2_T\, \sum^\infty_{j=0} c_j\, \alpha_t^j\,,~~ {\rm with} ~~ c_0 = 1\,,
\label{eq.kall.4}
\end{eqnarray}
where $\Lambda_T^2 = m^2_t\, \lambda^2_T$ is the Euclidean position of the tachyon in the on-shell scheme.
 It can be determined by solving numerically the following equation
\begin{eqnarray}
1 + \frac{1}{\lambda^2_T} = \alpha_t\, \log\big(\lambda_{T}^2\big)\,.
\label{eq.kall.5}
\end{eqnarray}
If we substitute the expansions in eqs.(\ref{eq.new.8}), (\ref{eq.new.13bis}) and (\ref{eq.kall.4}) 
 into eq.(\ref{eq.tac}) and we neglect the term $1/\lambda^2_{T,c}$, we get
\begin{eqnarray}
1 = \Big(\sum_{j=0} a_j\, \alpha_t^j\Big)\, 
\Big[1+\alpha_t\, \log\Big(\sum_{k=0} c_k\, \alpha_t^k\Big)+ \alpha_t\, W\Big]\,.
\label{eq.kall.6}
\end{eqnarray}
The above equation can be solved in the sense of formal power series, giving us the expansion coefficients 
 $\{ c_j\}$. We list here the first few of them.
\begin{eqnarray}
&&
c_1 = 0\,,\nonumber\\
&&
c_2 = \pi^2 \,,\nonumber\\
&&
c_3 = \frac{5}{2}\,\pi^2 \,,\nonumber\\
&&
c_4 = \frac{9}{2}\, \pi^2 \,.
\label{eq.kall.7}
\end{eqnarray}
It is interesting to notice that the expansion coefficients of $\lambda^2_{T,c}$ have the following remarkable 
property 
\begin{eqnarray}
\sum_{j=0}^n a_j\, c_{n-j} = 0\,, ~ ~ \forall n \geq 1\,.
\label{eq.kall.8}
\end{eqnarray}
The above relation entails that $\alpha_{t,c}\, \lambda^2_{T,c} = \alpha_t\, \lambda_T^2$ and consequently that 
 the position of the tachyon pole and its residuum are the same  
in the on-shell and in the complex mass scheme, i.e. $\Lambda_{T,c} \equiv \Lambda_T$ and $\kappa_c \equiv \kappa$.

%
\section{Perturbative top contributions to the $\rho$ parameter}
\label{sect.pert}
%

In this Section we shall use the resummed top propagator in eq.(\ref{eq.new.6}) in order to 
 compute perturbatively the leading contributions in the top quark mass to the $\rho$ parameter. 
We will also  show that the result one obtains in the framework of the complex mass scheme 
 coincides, upon rearranging the perturbative series in powers of $\alpha_t$, with the one in the  on-shell scheme.

At tree level $\rho = 1$ due to a global accidental symmetry, the so-called custodial symmetry. 
 At the leading order in the top quark mass, radiative corrections to $\rho$ stem from the transversal part 
 of the (unrenormalized) self-energies of the vector bosons $W$ and $Z$ in the low energy limit
\begin{eqnarray}
&&
\Delta \rho = \frac{\Pi_Z}{M_{Z}^2} -  \frac{\Pi_W}{M_{W}^2} \,\,,~ 
{\rm where~ with}\nonumber\\
&&
\Sigma^{\mu\nu}_V(p^2 = 0) = \Pi_V\, g^{\mu\nu}\,, V = W,Z\,\,,
\label{eq.58.3}
\end{eqnarray}
we denote  the vector self-energies at zero external momentum. 

Since the functional form of the resummed top propagator in the on-shell and in the complex mass scheme
 is the same, we can simply take the result obtained in Ref.~\cite{Bettinelli:2010gm} for $\Delta \rho$ 
\begin{eqnarray}
\Delta \rho = i\,\sqrt{2}\, N_c\, G_F\, m^4_{t,c}
\int \!\! \frac{d^4 q}{(2\pi)^4}\,
\frac{1}{\big[a_c(q^2) \,q^2-m^2_{t,c}\big]^2\,q^2}\,.
\label{eq.h12bis}
\end{eqnarray}
Although both IR- and UV-convergent, the above expression is ill-defined due to the presence of a tachyon pole. 
In particular the Wick rotation cannot be performed  because the resulting integral would be divergent.
However, since the tachyon is a non-perturbative effect, after expanding the denominator in eq.(\ref{eq.h12bis}) 
in powers of $\alpha_{t,c}$, we can perform the Wick rotation and the integration over the solid angle, finding 
\begin{eqnarray}
\Delta \rho = \frac{N_c}{N_F}\,\alpha_{t,c}\, \sum^{\infty}_{j=0}\big(j+1\big)\,\alpha_{t,c}^j
\,\int_{0}^{\infty} \!\! d x\,
\frac{x^j\,\big(\log x +W\big)^j}{\big(1+x\big)^{(j+2)}}\,,
\label{eq.n3}
\end{eqnarray}
where the dimensionless variable $x$ is given by $x = -\frac{q^2}{m^2_{t,c}}$.

All of the integrals in eq.(\ref{eq.n3}) can be computed by using the techniques 
developed in Ref.~\cite{Bettinelli:2010gm}. We give here only the final result.
\begin{eqnarray}
&&
\Delta \rho = \frac{N_c}{N_F}\,\sum^{\infty}_{j=0} r_j \,\alpha_{t,c}^{(j+1)} \,,~{\rm where}~~\nonumber\\
&&
r_j = \sum_{l =0}^{j} \sum_{k=0}^{\big[l/2\big]} 2 c_l(j)\,\frac{l!\,\, W^{(l-2k)}}{\big(l-2k\big)!}
\,\Big(1-2^{(1-2k)}\Big)\,\zeta\big(2k\big)\,.
\label{eq.n16bis}
\end{eqnarray}
The combinatorial coefficients $c_l(j)$ are related to the problem of grouping together 
 $l$ objects out of a total of $j$ without repetitions. Their explicit expression has been given 
 in Ref.~\cite{Bettinelli:2010gm}. Moreover  $\zeta(z)$ denotes the Riemann zeta function and finally $[\cdot]$ 
 is the integer part of a real number.
We report here the first few coefficients of the perturbative expansion of the $\rho$ parameter 
\begin{eqnarray}
&&\!\!\!\!\!\!\!\!\!
r_0 = 1\,, \nonumber\\
&&\!\!\!\!\!\!\!\!\!
r_1 = 1 + W\,,\nonumber\\
&&\!\!\!\!\!\!\!\!\!
r_2 = 1 + \frac{1}{3}\,\pi^2 + 3 W + W^2\,,\nonumber\\
&&\!\!\!\!\!\!\!\!\!
r_3 = 1 + \frac{11}{6}\,\pi^2  + \big(6 + \pi^2\big) W + \frac{11}{2}\,W^2 + W^3\,,\nonumber\\
&&\!\!\!\!\!\!\!\!\!
r_4 = 1 + \frac{35}{6}\,\pi^2 + \frac{7}{15}\,\pi^4 + \Big(10 + \frac{25}{3}\,\pi^2\Big) W +
      \Big(\frac{35}{2} + 2 \pi^2\Big) W^2 +\frac{25}{3}\, W^3 +W^4\,.
\label{eq.new.18}
\end{eqnarray}
Upon expanding the finite width effects, $W$, in powers of $\alpha_{t,c}$
\footnote{We remark that $W$ can be expressed as a formal power series in $\alpha_{t,c}$ first by solving 
 eq.(\ref{eq.new.7}) without expanding $\alpha_{t,c}$ in powers of $\alpha_t$ and then by plugging the solution 
 into the third line of eq.(\ref{eq.cms.6}). Since the technique has been already illustrated in Sec.~\ref{sec.top},
 we simply report here the final result
\begin{eqnarray*}
W = \pi^2\,\alpha_{t,c}\,\Big[1+\frac{3}{2}\,\alpha_{t,c}+\big(2+\pi^2\big) \alpha_{t,c}^2
+\Big(\frac{5}{2}+\frac{55}{12}\,\pi^2\Big)\alpha_{t,c}^3 + 
\big(3+13 \pi^2+2 \pi^4\big) \alpha_{t,c}^4 + O\big(\alpha_{t,c}^5\big) \Big]\,. 
\end{eqnarray*}
}, we get the perturbative expansion of $\Delta \rho$ in the framework of the complex mass scheme. 
\begin{eqnarray}
\Delta \rho \!\!\!&=&\!\!\! \frac{N_c}{N_F}\,\alpha_{t,c}\,\Big[1+\alpha_{t,c}+\Big(1+\frac{4}{3}\,\pi^2\Big)\alpha_{t,c}^2+
\Big(1+\frac{19}{3}\,\pi^2\Big)\alpha_{t,c}^3+\Big(1+\frac{55}{3}\,\pi^2+\frac{52}{15}\,\pi^4\Big)\alpha_{t,c}^4
\nonumber\\
&&~~~~~~~~\,\,\,
+\Big(1+\frac{125}{3}\,\pi^2+\frac{1406}{45}\,\pi^4\Big)\alpha_{t,c}^5+ O\big(\alpha_{t,c}^6\big)\Big].
\label{eq.exam.0}
\end{eqnarray}

It turns out that if we substitute the perturbative expansion of $\alpha_{t,c}$ in 
eq.(\ref{eq.new.8}) into the above equation and we reorder 
the series in powers of $\alpha_t$, we obtain exactly the same perturbative expansion
as in the on-shell scheme, namely    
\begin{eqnarray}
\Delta \rho \!\!\!&=&\!\!\! \frac{N_c}{N_F}\,\alpha_t\,\Big[1+\alpha_t+\Big(1+\frac{1}{3}\,\pi^2\Big)\alpha_t^2+
\Big(1+\frac{11}{6}\,\pi^2\Big)\alpha_t^3+\Big(1+\frac{35}{6}\,\pi^2
+\frac{7}{15}\,\pi^4\Big)\alpha_t^4\nonumber\\
&&~~~~~~~\,\,
+\Big(1+\frac{85}{6}\,\pi^2+\frac{959}{180}\,\pi^4\Big)\alpha_t^5+ O\big(\alpha_t^6\big)\Big].
\label{eq.exam.1}
\end{eqnarray}
This means that for an observable quantity, like the $\rho$ parameter, the two renormalization schemes 
 give identical results to all orders in perturbation theory. Moreover, we notice that, although both the perturbative
 expansions in the on-shell (\ref{eq.exam.1}) and in the complex mass scheme (\ref{eq.exam.0}) are divergent 
 asymptotic series, the coefficients of the former are systematically smaller.
This means that the perturbative expansion of the $\rho$ parameter, 
considered as a formal power series, has a better behaviour in the on shell scheme, 
since at fixed order the neglected terms are smaller.
Therefore the on shell mass appears to be more directly connected with the value of the radiative corrections. 
However, one cannot rigorously say it is more physical.
%

\section{Nonperturbative top contributions to the $\rho$ parameter}
\label{sect.npert}
%

In this Section we shall use the tachyon-free representation of the resummed top 
 propagator (\ref{eq.kall.3}) in order to compute nonperturbatively the exact leading top 
 quark contribution to the $\rho$ parameter at the leading order in the large $N_F$-limit.
It turns out that the result in the framework of the complex mass scheme coincides with the one 
in the on-shell scheme, proving in this way the independence of our procedure from the chosen 
 renormalization scheme. 

The contribution of the tachyonic subtraction term in eq.(\ref{eq.kall.3}) to the 
leading top contribution to $\Delta \rho$ can be computed following the procedure discussed at length 
 in Ref.~\cite{Bettinelli:2010gm}. We report here only the final result.
\begin{eqnarray}
\Delta \rho = \frac{N_c}{N_F}\,\alpha_{t,c}\,\frac{1}{\big(1-\kappa\big)^2}\,\int_0^{\infty} dx \,
\Bigg[\frac{1}{a_c(-x)\,x+1}+\frac{\kappa\,\lambda^2_{T,c}}{x-\lambda_{T,c}^2}\Bigg]^2\,.
\label{eq.fin.1}
\end{eqnarray}
After shifting the integration variable about the tachyon pole, $y := x- \lambda_{T,c}^2$, and using 
 the pole equation (\ref{eq.tac}), we get 
\begin{eqnarray}
&&\!\!\!\!\!\!\!\!\!\!\!\!\!\!\!\!\!\!\!\!\!
\Delta \rho = \frac{N_c}{N_F}\,\alpha_{t,c}\,\frac{1}{\big(1-\kappa\big)^2}\times \nonumber\\
&&\!\!\!\!\!
\int_{-\lambda^2_{T,c}}^{\infty} dy \,
\Bigg\{ \frac{1}{1-\Big(1+\frac{y}{\lambda_{T,c}^2}\Big)\Big[1+\alpha_{t,c}\,\lambda_{T,c}^2\,
\log\Big(1+\frac{y}{\lambda_{T,c}^2}\Big)\Big]}
+\frac{\kappa\,\lambda^2_{T,c}}{y}\Bigg\}^2\,.
\label{eq.fin.2}
\end{eqnarray}
Since the location of the tachyon does not depend on the renormalization scheme, one has 
\begin{eqnarray}
\lambda_{T,c}^2 = \frac{\lambda^2_T}{R}\,,~~ {\rm where}~~ 
R = \frac{m^2_{t,c}}{m^2_t} = \frac{\alpha_{t,c}}{\alpha_t}\,.
\label{eq.fin.3}
\end{eqnarray}
Finally if we rescale the integration variable, $\widetilde{y} := R\, y$, we immediately obtain
 the exact leading top contribution to the $\rho$ parameter in the on-shell scheme. This concludes the 
 proof of the equivalence of the two renormalization schemes.

\begin{figure}[t]
\begin{center}
\includegraphics[width=0.9\textwidth]{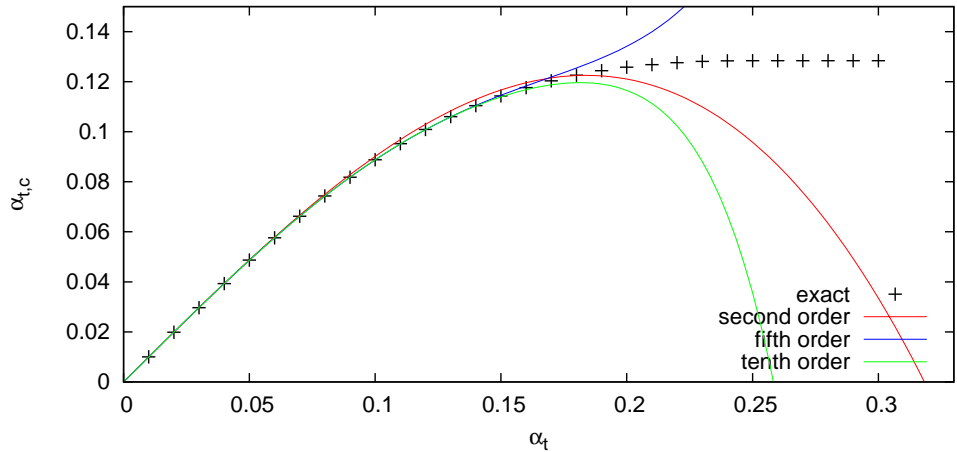}
\end{center}
\caption{Comparison between the exact behaviour of $\alpha_{t,c}$ and its perturbative expansion at the 
 second, fifth and tenth order}
\label{fig.1}
\end{figure}

It is interesting to determine the behaviour of the interaction strength of the theory in the complex 
 mass scheme, $\alpha_{t,c}$, as a function of the same quantity, but computed in the on-shell scheme, $\alpha_t$.
 First of all we solve numerically eq.(\ref{eq.new.7}). It turns out that this equation admits two solutions 
 for $0< \alpha_{t,c} < 0.128$, while no solution exists for bigger values of $\alpha_{t,c}$. The bigger solution 
 is unphysical because it diverges in the limit $\alpha_{t,c} \to 0$ and thus we discard it.
Then, the physical solution of eq.(\ref{eq.new.7}) can be used in order to get the finite width effects $W$.
 Finally, we compute the ratio $R$ by imposing the condition in eq.(\ref{eq.new.11})
\begin{eqnarray}
R-\alpha_{t,c}\,\log R +\alpha_{t,c}\, W - 1 = 0\,.
\label{eq.fin.4}
\end{eqnarray}
It turns out that the exact numerical result for $\alpha_{t,c}(\alpha_t)$ shows a linear dependence on $\alpha_t$ for
 $\alpha_t < 0.1$ and a typical saturation behaviour for bigger values of $\alpha_t$ (see Fig.~\ref{fig.1}). 
The perturbative expansion of $\alpha_{t,c}$ (\ref{eq.new.12}) nicely reproduces this behaviour due to the 
 fact that its terms have alternating signs.
%

\section{Naive fixed-width calculation}
\label{sec.fix}
%

While in our simplified case it is possible to calculate all
orders in perturbation theory, this is not generally possible
in more complicated cases like the Standard Model.
By necessity, one therefore often calculates with a fixed-width 
Breit-Wigner propagator within the loop. In this way, one of course gets the lowest
order correction correct, but one could ask the question whether
the contribution from the fixed-width captures a part of the higher-loop
corrections correctly at least approximately.
Within the fixed-width approximation the Fermi constants extracted from the low-energy 
limit of a neutral and a charged current process, i.e. $G_F^0$ and $G_F^+$ respectively, 
 become complex quantities. Thus, already at the one-loop level, 
one finds a complex $\rho$ parameter 
\begin{eqnarray} 
\rho_c = \frac{G_F^0}{G_F^+} = \frac{1}{1-\Delta \rho_c} = 1+ \frac{N_c}{N_F}\, \alpha_{t,c}\big(1-i w_t\big)\,.
\label{eq.BW.1}
\end{eqnarray}
What one measures is of course the absolute
value of the ratio $G_F^0/G_F^+$ and one defines $\Delta \rho$ on this basis.
Using this naive prescription we find the following formula
for the perturbative expansion of $\Delta \rho$ in powers of $\alpha_t$:
\begin{eqnarray}
\Delta \rho =\frac{N_c}{N_F}\,\alpha_t\,\Big[1-\pi^2 \alpha_t^2-
\frac{5}{2}\,\pi^2 \alpha_t^3 + O\big(\alpha_t^4\big)\Big]\,. 
\label{eq.BW.2}
\end{eqnarray}
From the above equation, we see that the coefficients do not match even approximately
the exact result (\ref{eq.exam.1}). The subleading term of $O\big(\alpha_t^2\big)$ is actually absent.
We conclude therefore, that there is no shortcut to estimate the effects
of higher orders. One really has to calculate them.

\section{Conclusions}
\label{sect.con}
%

In this Letter we considered the $SU(N_F) \times U(1)$ model at the leading order in the large 
 $N_F$-limit in the framework of the complex mass scheme. We computed the exact leading top 
 quark contribution to the $\rho$ parameter and its perturbative expansion to all orders in 
 perturbation theory, showing that they coincide with the same results obtained by adopting the 
 on-shell renormalization scheme.

At the perturbative level, a comparison between the two renormalization schemes is obtained by expanding 
 the complex subtraction point, $\mu_t$, in powers of the interaction strength of the theory in 
 the on-shell scheme, $\alpha_t$. This expansion allows us to convert the perturbative definition of any 
 observable quantity in the complex mass scheme to the on-shell scheme. 

In order to go beyond the perturbative approximation, one has to take care of the tachyonic pole which 
 is present in the exact top propagator. It turned out that both the Euclidean position of the tachyon 
 and its residuum do not depend on the chosen renormalization scheme. We regularized the resummed  
 propagator by subtracting the tachyon minimally at its pole. Although not unique, this procedure allows 
 to define a tachyon-free representation of the exact top propagator which respects gauge invariance.
  The latter has been used to determine an expression (\ref{eq.fin.1}) for the exact leading 
top contribution to the $\rho$ parameter which can be evaluated numerically. 
The exact numerical results in the two renormalization schemes 
considered coincide, being related by a change of the Feynman-like integration variable. 

We showed that a naive treatment with a fixed width for the top quark cannot give even 
approximately correct results.

\section*{Acknowledgements}
We would like to thank O. Brein for a critical reading of the manuscript.
This work is supported by the DFG project "{(Nicht)-perturbative Quantenfeldtheorie\,}".

\end{document}